\documentclass[aip,jcp,amsmath,amssymb,reprint,twocolumn]{revtex4-1}
\usepackage{graphicx}
\usepackage{hyperref}
\usepackage{dcolumn}                 
\usepackage{times}
\usepackage[utf8]{inputenc}
\usepackage[T1]{fontenc}
\newcolumntype{L}{>{$}l<{$}}

\newcommand*{\cent}[1]{\multicolumn{1}{c}{$#1$}}

\begin{document}
\preprint{Version 2.0}

\title{Long-range asymptotics of exchange energy in the hydrogen molecule}

\author{Micha{\l} Si{\l}kowski}
\email{michal.silkowski@fuw.edu.pl}
\author{Krzysztof Pachucki}
\affiliation{Faculty of Physics, University of Warsaw, Pasteura 5, PL 02-093 Warsaw, Poland} 

\begin{abstract}
The exchange energy, i.e. the splitting $\Delta E$ between gerade and ungerade states in the hydrogen
molecule has proven very difficult in numerical calculation at large internuclear distances $R$,
while known results are sparse and highly inaccurate. On the other hand,
there are conflicting analytical results in the literature concerning its asymptotics.
In this work we develop a flexible and efficient numerical approach using explicitly correlated exponential functions
and demonstrate highly accurate exchange energies for internuclear distances as large as 57.5 au.
This approach may find further applications in calculations of inter-atomic interactions. In particular,
our results support the asymptotics form $\Delta E \sim R^{5/2}e^{-2R}$,
but with the leading coefficient being $2\,\sigma$ away from the analytically derived value. 
\end{abstract}

\maketitle

\section{Introduction}
The calculations of molecular properties like rovibrational energies and various relativistic effects
are routinely performed at small internuclear distances $R$.
Numerical results with standard available quantum mechanical codes
are not guaranteed to be accurate for large $R$, especially for exponentially small quantities.
While the use of exponential functions with explicit electron correlations has become the natural choice
for an approach with correct wave function asymptotics,
no efficient method to perform corresponding integrals has been developed.
Namely, the original calculations of Ko{\l}os and Roothaan \cite{kolos1} with exponential functions
were an extension of earlier developments by Podolanski \cite{podolanski} and R\"udenberg \cite{ruedenberg}.
Their method was based on Neumann expansion of the $1/r_{12}$ term in spheroidal coordinates,
which has a very slow numerical convergence.
In spite of this, a few years later Ko{\l}os and Wolniewicz \cite{kolos2} were able to obtain very accurate, for the
time, Born-Oppenheimer energies. This was a remarkable breakthrough in the field of precision molecular structure.
Thus far, however, no accurate results for exchange energy have been obtained for large internuclear
distances. For this reason the discrepancies between
conflicting analytic results have not yet been resolved.

Before going into detail, let us introduce the notion of the exchange energy.
The hydrogen molecule, assuming clamped nuclei, is described by the electronic wave function, which can be symmetric
or antisymmetric with respect to the exchange of electrons and with respect to inversion through the geometrical center.
For a large internuclear separation the symmetry of the wave function does not matter, and one has two separate hydrogen atoms.
It means that the difference between symmetric and antisymmetric state energies has to be exponentially small.
Indeed, this splitting behaves as $e^{-2 R}$, but the question remains concerning the prefactor.
If we assume that the wave function is just a product of two properly symmetrized hydrogen orbitals,
the so-called Heitler-London wave function~\cite{heitler}, then the splitting is of the form given by Eq.\,(\ref{hl}),
which is known to be invalid, even with regard to its sign~\cite{critique}.
It means that, even for large internuclear distances, one cannot assume that the electronic wave function is a symmetrized product
of hydrogenic ones. Therefore, the  asymptotics of the exchange energy
and the functional form of the wave function that reproduces this asymptotic behavior is an interesting problem.\\

There are conflicting results among analytic calculations of the exchange energy for the leading term,
and there are no conclusive results for subleading ones. The problem seems to be so difficult that no successful attempts
to resolve these discrepancies have been reported so far. In this work, 
we develop a numerical approach to calculate the exchange energy at large internuclear distances
with well controlled numerical accuracy.
To achieve this, we employed a large basis of exponential functions of the generalized Heitler-London form with
an arbitrary polynomial in all interparticle distances. We have previously developed an efficient recursive approach
to calculate integrals with two-center exponential functions~\cite{kw},
and demonstrated its advantages by the most accurate calculation of Born-Oppenheimer (BO) energies for the hydrogen molecule.
In this work we use an efficient parallel version of linear algebra \cite{hsl,hogg} in an arbitrary precision arithmetic
to fully control the precision of the exchange energy.
For example, about 230-digit arithmetic is utilized at the largest internuclear separations
of $57.5$ au to obtain six significant digits of the exchange energy out of 50 digits for the total energy.
To our knowledge, there is no better approach available that will give the exchange energy with full
control of the numerical precision.
In fact, the widely used Symmetry Adapted Perturbation Theory (SAPT) method~\cite{sapt}, which aims to extract the exchange energy
contribution in a perturbative manner, has pathologically slow convergence~\cite{sapt2,sapt3} for large internuclear
distances, and its results are far from accurate.

\section{Formulation of the problem}

Let us consider a stationary Schr\"odinger equation for two electrons with positions given by $\vec r_1$ and $\vec r_2$ in the H$_2$ molecule
in the BO approximation with internuclear separation $R$,
\begin{equation}
	H\Psi_g(\vec r_1\,,\vec r_2) = E_g(R)\Psi_g(\vec r_1\,,\vec r_2),
\end{equation}
\begin{equation}
	H\Psi_u(\vec r_1\,,\vec r_2) = E_u(R)\Psi_u(\vec r_1\,,\vec r_2),
\end{equation}
where subscripts $g$ and $u$ denote \emph{gerade} and \emph{ungerade} symmetry under the inversion with respect to the
geometrical center. They correspond to the ground electronic states with a total spin $S=0$ and $S=1$. 
$H$ in the above equation is the nonrelativistic Hamiltonian of the hydrogen molecule with clamped nuclei,
\begin{equation}
H =
-\frac{\nabla_1^2}{2}-\frac{\nabla_2^2}{2}-\frac{1}{r_{1A}}-\frac{1}{r_{1B}}
-\frac{1}{r_{2A}}-\frac{1}{r_{2B}} +\frac{1}{r_{12}}
+\frac{1}{R}.
\end{equation}
Within the BO approximation, one considers only the electronic part of the wave function
of the system, and thus $R$ serves as a parameter for electronic energies.
The difference between these energies
\begin{align}
  \Delta E = E_{\rm u} - E_{\rm g}
\end{align}
is the \emph{energy splitting}, and  half of this splitting with the minus sign, $J = -\Delta E/2$,
was the exchange energy according to the definition used in previous works. In this work we always refer to $\Delta E$
and convert results of the previous works from $J$ to $\Delta E$. Consequently, we use the exchange energy as a synonym of the energy splitting.

The pioneering theory of Heitler and London~\cite{heitler} was one of the first attempts
to explain chemical bonding~\cite{valence} on the grounds of the freshly
established foundations of quantum mechanics.
Their method was based on approximation of the wave functions corresponding to the lowest energy states of
H$_2$ with symmetrized and antisymmetrized products of the exact hydrogen atom solutions.
This approach was pursued in the same year by Sugiura~\cite{sugiura} who derived Born-Oppenheimer energies of the lowest
gerade and ungerade states as a function of $R$. The asymptotic value for the energy splitting based on Sugiura~\cite{sugiura} reads,
\begin{eqnarray}\label{hl}
	\Delta E_{\rm HL}(R) &=& 2\left\lbrack \frac{28}{45} - \frac{2}{15}\left(\ln R + \gamma_E \right) \right\rbrack
	R^3\exp(-2R)\nonumber \\ &&
+ \mathcal{O}\left(R^2\exp(-2R)\right),
\end{eqnarray}
where $\gamma_E = 0.577~215$... is the Euler-Mascheroni constant.

The Heitler-London approach appeared plausible because it provided a reasonable mechanism of chemical bond formation and its
comparison with the results of ab-initio numerical calculations~\cite{kolos1} obtained later was satisfactory.
Nevertheless, its long-range asymptotics is inherently flawed based on mathematical grounds. Analysis of Eq.\,(\ref{hl})
reveals that the asymptotic behavior of $\Delta E$ in Heitler-London theory becomes unacceptable,
due to the logarithmic term being dominant as $R\rightarrow \infty$. As a consequence, for sufficiently large $R$
($\approx 60$ au)
the energy splitting becomes negative. The negative sign of $\Delta E_{\rm HL}(R)$ contradicts the well-established theorem
on Sturm-Liouville operators with homogeneous boundary conditions stating that the lowest energy eigenstate
should be nodeless in the coordinate space.

Many years later, it was recognized that the Heitler-London approach underestimates electron correlations~\cite{critique},
and the logarithmic term in Eq.\,(\ref{hl}) could be identified with a potential coming from
the \emph{exchange} charge distribution.
This line of reasoning led to conjectures (see Refs.~\cite{tang,tang2,burrows}) that the correct asymptotics
might be in the same form as Heitler-London if only the logarithmic term is appropriately suppressed
(e.g. via proper treatment of electron correlations).
Indeed, Burrows \emph{et al.}~\cite{burrows}, on the grounds of algebraic perturbation
theory~\cite{burrows2}, have derived their formula for the long-range asymptotics of the splitting
\begin{equation}\label{bdc}
	\Delta E_{\rm BDC}(R) = R^3e^{-2R}\left(\gamma_{\rm BDC}+ \mathcal{O}\left(\frac{1}{R}\right) \right),
\end{equation}
with $\gamma_{\rm BDC}=0.301\,672$..., a result similar to the Heitler-London result given by Eq.\,(\ref{hl})
aside from the unphysical logarithmic term.

A completely different line of reasoning was introduced by Gor'kov and Pitaevskii~\cite{gorkov}, followed only a few
months later by a very similar method by Herring and Flicker~\cite{herring}.
They both used a kind of quasiclassical approximation for the wave function to derive its asymptotic form,
and with the help of a surface integral summarized by Eq.\,(\ref{sim}), they obtained the exchange energy.
A mistake in the numerical coefficient of the leading term in the former was indicated and corrected
in the latter paper~\cite{herring}, and their final result is 
\begin{equation}\label{hf}
	\Delta E_{\rm HF} = \gamma_{\rm HF} R^{5/2}\exp(-2R) + \mathcal{O}\left(R^2\exp(-2R)\right),
\end{equation}
with the leading order coefficient
\begin{align}
	\gamma_{\rm HF} = 1.636\,572\,063\ldots \label{gammaHF}
\end{align}  
Accounting for the asymptotic wave function requires careful analysis of various regions of the 6-dimensional
space (see for instance Ref.~\cite{herring}).
With the increasing internuclear distance the exact wave function approaches a symmetrized or
antisymmetrized product of two isolated hydrogen atom solutions.
Nonetheless, the exact way in which this limit is approached is of paramount importance for the asymptotics of
exchange energy, as has been thoroughly discussed with the case of Heitler-London theory in Ref.~\cite{critique}.
In comparison to analytic approaches for H$_2^+$~\cite{damburg,cizek,gniewek,gniewek2}, examination of asymptotic energy
splitting in H$_2$ is substantially more challenging due to electron-electron correlation,
and it is prone to mistakes, as made evident by the presence of conflicting results in
the literature~\cite{herring,gorkov,tang2,burrows}, compare Eqs. (\ref{bdc}) and (\ref{hf}).

\section{Derivation of the leading asymptotics}

To our knowledge, all of the analytic derivations presented in Refs.~\cite{gorkov,herring,andreev,tang,tang2,scott,burrows,burrows2} ultimately
rely on the Surface Integral Method (SIM), also referred to in the literature as the
Smirnov~\cite{smirnov} or Holstein-Herring~\cite{holstein} method.
Here we follow the work of Gor'kov and Pitaevskii~\cite{gorkov} to present
the SIM and the derivation of the asymptotic exchange energy in Eq.\,(\ref{hf}).
This derivation lacks mathematical rigor but nevertheless
helped us to understand the crucial behavior of the asymptotic wave function.
Moreover, their asymptotics is confirmed by our numerical calculations, which is only
twice the uncertainty ($2\,\sigma$) away from their analytical value.

Let us assume that nuclei are on the $z$-axis with $z=a,-a$, $(R=2\,a)$, and
let $\Omega$ be half of the 6-dimensional space with $z_2\geq z_1$, and $\Sigma$ is a boundary of $\Omega$,
namely 5-dimensional space with $z_1=z_2$. Consider the following integral
\begin{align}\label{sim}
  &\ (E_g-E_u)\int_\Omega d^3r_1\,d^3r_2\,\Psi_g\,\Psi_u \nonumber \\ =&\
  \frac{1}{2}\,\int_\Omega d^3r_1\,d^3r_2\, \bigl[\Psi_g\,(\Delta_1+\Delta_2)\,\Psi_u - \Psi_u\,(\Delta_1+\Delta_2)\,\Psi_g\bigr]
  \nonumber \\ =&\
  \int_\Omega d^3r_1\,d^3r_2\, \vec\nabla_1\bigl[\Psi_g\,\vec\nabla_1\,\Psi_u - \Psi_u\,\vec\nabla_1\,\Psi_g\bigr]
  \nonumber \\ =&\
  \oint_\Sigma d\vec S\,\bigl[\Psi_g\,\vec\nabla_1\,\Psi_u - \Psi_u\,\vec\nabla_1\,\Psi_g\bigr].
\end{align}
which allows us to express the energy splitting in terms of a surface integral with $\Psi_g$ and $\Psi_u$ functions.
Let us introduce a combination of these functions
\begin{align}
  \Psi_1 =&\ \frac{1}{\sqrt{2}}(\Psi_g+\Psi_u),\\
  \Psi_2 =&\ \frac{1}{\sqrt{2}}(\Psi_g-\Psi_u),
\end{align}
with the respective phase chosen in a manner such that $\Psi_{1,2}$ are real and correspond
to an electron localized at a specific nucleus, namely
\begin{align}
\Psi_1 \approx \frac{1}{\pi}\,e^{-|\vec r_1+\vec a| -|\vec r_2-\vec a|},\;\;\mbox{\rm for}\; \vec r_1\rightarrow -\vec a;\;\vec r_2\rightarrow \vec a\\
\Psi_2 \approx \frac{1}{\pi}\,e^{-|\vec r_1-\vec a| -|\vec r_2+\vec a|},\;\;\mbox{\rm for}\; \vec r_1\rightarrow \vec a;\;\vec r_2\rightarrow -\vec a
\end{align}
and the $\Psi_{g/u}$ functions are normalized to 1. The left-hand side of Eq.\,(\ref{sim}) can be transformed to
\begin{align}
  \int_\Omega \! d^3r_1\,d^3r_2\,\Psi_g \Psi_u
  =&\ \frac{1}{2}\int_\Omega d^3r_1\,d^3r_2 \bigl[\Psi_g^2+\Psi_u^2 - (\Psi_g-\Psi_u)^2\bigr]\nonumber \\
  =&\ \frac{1}{2}-\int_\Omega d^3r_1\,d^3r_2\,\Psi_2^2
\end{align}
and the right hand side to
\begin{align}
 &\ \oint_\Sigma d\vec S\,\bigl[\Psi_g\,\vec\nabla_1\,\Psi_u - \Psi_u\,\vec\nabla_1\,\Psi_g\bigr] \nonumber \\ =&\ 
\oint_\Sigma d\vec S\,\bigl[\Psi_2\,\vec\nabla_1\,\Psi_1 - \Psi_1\,\vec\nabla_1\,\Psi_2\bigr].
\end{align}
As a result, one obtains
\begin{align}
  E_g-E_u = \frac{2\,\oint_\Sigma d\vec S\,\bigl[\Psi_2\,\vec\nabla_1\,\Psi_1 - \Psi_1\,\vec\nabla_1\,\Psi_2\bigr]}
                 {1-2\,\int_\Omega dV\,\Psi_2^2}. \label{surfaceint}
\end{align}
The second term in the denominator is exponentially small, and thus can be safely neglected.
By virtue of Eq.\,(\ref{surfaceint}), the knowledge of the wave function and its derivative on $\Sigma$ is sufficient to retrieve the energy splitting.
The advantage of this method manifests itself especially in the regime of large internuclear distance, where exact
wave functions of singlet and triplet states are close to the appropriately symmetrized and antisymmetrized products
of the isolated hydrogen atom solutions.

Below we closely follow the procedure of Ref.~\cite{gorkov} and correct several misprints there.
Let us assume the following ansatz of the wave functions $\Psi_{1/2}$
\begin{align}
  \Psi_1(\vec r_1,\vec r_2) =&\ \frac{\chi_1(\vec r_1,\vec r_2)}{\pi}\, e^{-|\vec r_1+\vec a| -|\vec r_2-\vec a|},\\
  \Psi_2(\vec r_1,\vec r_2) =&\ \frac{\chi_2(\vec r_1,\vec r_2)}{\pi}\, e^{-|\vec r_1-\vec a| -|\vec r_2+\vec a|},
  \end{align}
where the functions $\chi_{1/2}$ change slowly in comparison to the exponential terms.
Let us consider the region of $z_1\approx a, z_2\approx -a$ and $\rho_1,\rho_2\sim\sqrt{a}$ where the exponentials become
\begin{align}
  &\ e^{-|\vec r_1+\vec a| -|\vec r_2-\vec a|} \nonumber \\ &\
  \sim \exp\Bigl\{-2\,a-z_1+z_2-\frac{\rho_1^2}{2\,(a+z_1)} - \frac{\rho_2^2}{2\,(a-z_2)}\Bigr\},
  \end{align}
and $\rho_i$ is the perpendicular distance of $i$-th electron from the internuclear axis. From the Schr\"odinger equation one obtains for $\chi_1$
\begin{align}
	\label{chi1}
  \biggl[\frac{\partial}{\partial z_1} - \frac{\partial}{\partial z_2} + \frac{1}{2\,a}-\frac{1}{a-z_1}
-\frac{1}{a+z_2} + \frac{1}{|\vec r_1-\vec r_2|} \biggr]\,\chi_1 = 0,
\end{align}    
where higher order $O(1/\sqrt{a})$ terms are neglected.
Introducing $z_1 = (\xi+\eta)/2$ and $z_2 = (\xi-\eta)/2$, this equation takes the form
\begin{align}
	\label{chi1neg}
  \biggl[\frac{\partial}{\partial\eta} -\frac{1}{2\,a-\xi-\eta} - \frac{1}{2\,a+\xi-\eta}
    +\frac{1}{2\,\sqrt{\eta^2+\rho_{12}^2}} + \frac{1}{4\,a} \biggr]\chi_1 = 0,
  \end{align}
and the general solution is
\begin{align}
  \chi_1 =&\ C(\xi,\rho_{12})\,e^{-\frac{\eta}{4\,a}}\,\frac{\sqrt{\sqrt{\eta^2+\rho_{12}^2}-\eta}}
                                 {(2\,a-\xi-\eta)\,(2\,a+\xi-\eta)},
\end{align}  
up to the unknown function $C(\xi,\rho_{12})$. This function, which is not a rigorous argument,
is determined from the condition that whenever $\vec r_1\approx -\vec a$ or  $\vec r_2\approx \vec a$ the wave function $\Psi_1$
should be just exponential, and thus $\chi_1=1$ in this region. This argument can be justified
because for $\vec r_1\approx -\vec a$, the second electron interacts dominantly with its nucleus.
From this condition one obtains
\begin{widetext}
  \begin{align}
    \Psi_1(\vec r_1,\vec r_2) =&\ \frac{2\,a\,(2\,a-|z_1+z_2|)}{\pi\,(a-z_1)(a+z_1)}\,
    \exp\biggl(-2\,a-z_1+z_2 - \frac{\rho_1^2}{2\,(a+z_1)} - \frac{\rho_2^2}{2\,(a-z_2)}\biggr)\nonumber \\ &\
    \times\sqrt{\frac{\sqrt{(z_1-z_2)^2+\rho_{12}^2}+z_2-z_1} {\sqrt{(2\,a -|z_1+z_2|)^2+\rho_{12}^2}+2\,a-|z_1+z_2|}}
\exp\biggl(-\frac{1}{2} +\frac{z_2-z_1+|z_1+z_2|}{4\,a}\biggr).
  \end{align}  
  \end{widetext}
The function $\Psi_2$ is obtained by the replacement $\vec r_1\leftrightarrow \vec r_2$.
The appearance of $\rho_{12}$ in the  wave functions $\Psi_1$ is in crucial distinction to the Heitler-London wave function
and ensures the correct sign of the leading order asymptotics for all distances, as pointed out in Ref.~\cite{umanski}.
$\Psi_1$, however, is not differentiable at $z_1+z_2=0$ and this is one of the reasons we were not able to fully accept
this derivation. A similar problem appears in a later derivation of Herring and Flicker~\cite{herring}
and this lack of analyticity at $z_1+z_2=0$ was somehow ignored in all the previous works.
One may even ask, why this nonanalytic wave function should give the right asymptotics,
and here we show that, indeed, $\gamma\,R^{5/2}\,e^{-2\,R}$ behavior is in agreement with our numerical calculations,
although $\gamma$ is $2\sigma$ away.

Let us now return to Eq.\,(\ref{surfaceint}) to obtain the energy splitting from the above $\Psi_i$.
Because $\chi_{1/2}$ is slowly changing in comparison to dominant exponentials,
their derivative can be neglected, and the splitting becomes
\begin{align}\label{gpint}
  E_g-E_u =&\ -8\,\int_0^a dz\,d^2\rho_1d^2\rho_2\;\Psi_2\,\Psi_1\biggr|_{z_1=z_2=z},
\end{align}
where 
\begin{align}
	\Psi_2\,\Psi_1\biggr|_{z_1=z_2=z>0} =
	\frac{(2a)^2\,\rho_{12}\,e^{-4a-1} }{\pi^2(a-z)(a+z)^2}\,e^{\frac{z}{a}+\frac{a(\rho_1^2+\rho_2^2)}{z^2-a^2}},
\end{align}
where only the leading terms in the limit of a large $a$ are retained.
Integrals over $\vec \rho_1$, $\vec \rho_2$  yield
\begin{align}
	\int \mathrm{d}^2\rho_1 \mathrm{d}^2\rho_2\;\rho_{12} e^{-\alpha(\rho_1^2+\rho_2^2)} =&\ \pi^2\sqrt{\frac{\pi}{2}}
	\alpha^{-5/2}.
\end{align}
As a result one is left with the one-dimensional $z$-integral
\begin{align}
	E_g-E_u = -32 e^{-4a-1} \sqrt{\frac{\pi}{2a}}\int_0^a \! dz\, e^{z/a} (a-z)^{3/2}(a+z)^{1/2},
\end{align}
which after change of a variable $q=1-z/a$ yields
\begin{align}
	E_g-E_u =&\ -16\sqrt{2\pi} a^{5/2} e^{-4a} \,\int_0^1 dq\, e^{-q} q^{3/2}(2-q)^{1/2}.
\end{align}
After noting that $a=\frac{R}{2}$, this result 
\begin{align}\label{gamma}
  \gamma =4\,\sqrt{\pi}\,\int_0^1dq\,e^{-q}\,q^{3/2}\,(2-q)^{1/2} 
\end{align}
coincides with Eq.~(19) of Ref.~\cite{herring}.
The numerical value $\gamma = \gamma_{\rm HF}$ from Eq. (\ref{gammaHF})
will be verified in the next Sections by direct numerical calculations of the exchange energy.

\section{Variational approach}

In the simplest implementation of the variational approach one solves the Schr\"odinger equation by representing the
wave function as a linear combination of some basis functions and finds linear coefficients without optimization of nonlinear
ones. Because we are interested in the large $R$ asymptotics
of the exchange energy, the only viable option is to employ an exponential basis. This ensures
the short-range \emph{cusp conditions}~\cite{kato} and correct long-range asymptotic behavior
of the trial wave function. Consequently, the basis of trial functions is chosen in the following form
\begin{eqnarray}
	\phi &=& \sum_{\{n_i\}}\,c_{\{n_i\}}(1\pm P_{AB})\,(1\pm P_{12})
	\,e^{-r_{1A}-r_{2B}}\nonumber \\ &&
	\times r_{12}^{n_1}\,\eta_1^{n_2}\,\eta_2^{n_3}\,\xi_1^{n_4}\,\xi_2^{n_5},
\end{eqnarray}
where
\begin{equation}
	\eta_i = r_{iA}-r_{iB},~\xi_i = r_{iA}+r_{iB},
\end{equation}
and where $P_{AB}$ and $P_{12}$ represent operators enforcing symmetry with respect to the permutation $r_A \leftrightarrow r_B$
and $r_1 \leftrightarrow r_2$.
Only one type of exponent is used in the wave function, because the ionic structures like H${^+}$H${^-}$,
which correspond to a different choice of the exponent $e^{-r_{1A}-r_{1B}}$,
are subdominant in our problem, as was already discussed in Ref.~\cite{critique}, and thus they can be omitted.

It is tempting to assume that the sum over non-negative integer indices ${n_i}$ is chosen such
that for the so-called shell number $\Omega$ 
\begin{equation}\label{shell}
	\sum_{i=1}^5 n_i \le \Omega,
\end{equation}
because it gives a good numerical convergence for the total binding energy.
However, the main problem here is the very low numerical convergence of the exchange energy
at large internuclear distances $R$ with the increasing size of the basis as given by $\Omega$.
It was the main reason that previous numerical attempts were not very successful.

We notice that for $R\to\infty$, the main contribution to the numerator of the surface integral in Eq.\,(\ref{surfaceint})
comes from the integration over the neighborhood of the internuclear axis. We thus anticipate that the crucial
behavior of the wave function is encoded in $\eta_{1,2}$.
Consequently, a basis is constructed using three independent shell parameters,
such that the sum of powers of $\eta_{1,2}$, $\xi_{1,2}$, and $r_{12}$ are controlled by corresponding shell
numbers $\Omega_A$, $\Omega_B$, and $\Omega_C$
\begin{equation}\label{shellm}
n_2+n_3 \le \Omega_A,~n_4+n_5 \le \Omega_B,~n_1 \le \Omega_C,
\end{equation}
and numerical convergence is attained independently in each shell parameter.

Matrix elements of the nonrelativistic Hamiltonian in this basis can be expressed in terms of direct and
exchange integrals of the form 
\begin{align}
	f_{\{n_i\}}(R) =&\ R\!\int\! \frac{d^3 r_1}{4\,\pi}\!\int\! \frac{d^3 r_2}{4\,\pi}
	\frac{e^{-w_1\,r_{12}-y\,\eta_1-x\,\eta_2-u\,\xi_1-w\,\xi_2}}{r_{1A}\,r_{1B}\,r_{2A}\,r_{2B}}\nonumber \\ &\ 
\times r_{12}^{(n_1-1)}\,\eta_1^{n_2}\,\eta_2^{n_3}\,\xi_1^{n_4}\,\xi_2^{n_5},
\end{align}
with non-negative integers $n_i$.
When all $n_i=0$ $f$ is called the master integral, see Ref.~\cite{kw}
\begin{align}
	f(R) =&\ R \int \frac{d^3 r_1}{4\,\pi}\,\int \frac{d^3 r_2}{4\,\pi}\,
	\frac{e^{-w_1\,r_{12}-y\,\eta_1-x\,\eta_2-u\,\xi_1-w\,\xi_2}}{r_{12}~r_{1A}~r_{2A}~r_{1B}~r_{2B}}.
\end{align}
All integrals $f_{n_1n_2n_3n_4n_5}(r)$ can be constructed through differentiation of
the master integral with respect to the nonlinear parameters
and can be reformulated into stable recurrence relations~\cite{kw}, providing a way
to obtain all the integrals required to build matrix elements.
Details on the computation of necessary integrals and matrix elements can be found in our previous works in
Refs.~\cite{h2solv,rec_h2,kw}.

Having constructed the Hamiltonian and overlap matrices,
the energy and linear coefficients $c_{n_0 \ldots n_4}$ are determined by the secular equation,
\begin{equation}\label{eig}
	\det\left\lbrack \left\langle n_1 \ldots n_5 \big\vert H \big\vert n^{\prime}_1 \ldots n^{\prime}_5 \right\rangle
	- E \left\langle n_1 \ldots n_5 \big\vert n^{\prime}_1 \ldots n^{\prime}_5 \right\rangle \right\rbrack = 0.
\end{equation}
It has to be solved separately for $E_g$ and $E_u$. Consequently, to retrieve
an exponentially small difference between eigenvalues for large internuclear distances,
employment of extended-precision arithmetic is inevitable.

The generalized eigenproblem in Eq.\,(\ref{eig}) is solved with the help of the Shifted Inverse Power Method.
At each iteration the linear system has to be solved to refine the initial eigenvalue estimation,
which is done via calculation of the exact Cholesky factor of the $H-ES$ matrix, where $H$ is the Hamiltonian and $S$ the overlap matrix.
A significant drawback of the applied basis is the fact that those matrices are dense, far from diagonally dominant and
near-singular, especially for large $R$. 
This specific structure of matrices in the explicitly correlated exponential basis does not allow for straightforward application
of iterative (e.g. Krylov-like) methods.
Computation of the exact inverse Cholesky factor proved to be a suitable approach, providing cubic convergence.
A crucial advantage of this method is that the Cholesky factor has to be computed only once and can be reused in every iteration.
The main drawback of performing full Cholesky factorization is its algorithmic complexity.
It requires $n^3/3$ arithmetic operations in arbitrary precision,
which eventually become a bottleneck of the whole calculation. Total computation time can be significantly reduced
when Cholesky factorization is parallelized. We found that our implementation of procedure HSL\_MP54
for dense Cholesky factorization from the HSL library~\cite{hsl,hogg} adopted to
arbitrary precision performed best in terms of performance and accuracy.

Relying on our previous calculations of the Born-Oppenheimer potential for H$_2$~\cite{bo_h2} and anticipating that
variational calculations will follow one of the analytic results for the leading asymptotics of energy splitting, $\Delta
E_{\rm BCD}(R) \sim R^3e^{-2R}$ or $\Delta E_{\rm HF}(R) \sim R^{5/2}e^{-2R}$ with the coefficient of order of unity, the
accuracy goal in decimal digits can be estimated as $\lceil\log_{10}\left(\Delta E\right)\rceil + n$.
It is the number of correct digits in $E_g$ and $E_u$ required to obtain the difference between them on $n$ last significant digits.
In the extreme case of the internuclear distance  $R=57.5$ au, approximately 50 correct digits in the final numerical value of
$E_g$ and $E_u$ are required. Consequently, solving generalized eigenvalue problems for $E_g$ and $E_u$ requires
incorporation of arbitrary precision software. We have found the MPFR library~\cite{mpfr} to be robust and
provide the best performance among all the publicly available arbitrary-precision software.

\section{Numerical results} \label{results}
\begin{table*}[t]\renewcommand{\arraystretch}{1.2}
  \caption{\label{table1} Dependence of energy splitting $\Delta E = E_u-E_g$ scaled by
factor $R^{-5/2}e^{2R}$ on shell parameters at different
internuclear distances $R$ [au]. The first column presents $\Delta E$ as obtained with $\Omega_B=0,\Omega_C=0$, i.e. with
no explicit correlation in the basis. The second is $\Delta$, the difference in energy splitting between
$\Omega_B=\Omega_C=0$ basis and $\Omega_C\ne0$ basis, and the third column is its value as extrapolated in
$\Omega_C$, still with $\Omega_B$ fixed at zero. The fourth column shows a correction $\delta$ to $\Delta E$
due to $\Omega_B\neq 0$, and the last column is the total energy splitting. 
Uncertainty of $\Delta E(\Omega_B=0,\Omega_C=0)$, $\Delta$ and $\delta$ come from extrapolation in $\Omega_A$,
$\Omega_C$, and $\Omega_B$, respectively, and were obtained as described in the text.}
\centering
\begin{ruledtabular}
  \begin{tabular}{cLLLLL}
	  \cent{R} & \cent{\Delta E(\Omega_B=0,\Omega_C=0)} & \cent{\Delta} & \cent{\Delta E(\Omega_B=0)} & \cent{\delta} & \cent{\Delta E} \\
	\hline
	20.0 & 1.418\,595\,21(9) & 0.138\,969\,8(18) & 1.557\,565\,0(18) & -0.006\,16(27)  & 1.551\,41(27) \\
	22.5 & 1.409\,067\,90(8) & 0.140\,756\,3(20) & 1.549\,824\,2(20) & -0.004\,76(22)  & 1.545\,06(22) \\
	25.0 & 1.402\,295\,96(8) & 0.142\,648\,3(22) & 1.544\,944\,3(22) & -0.003\,69(18)  & 1.541\,25(18) \\
	27.5 & 1.397\,382\,94(7) & 0.144\,560\,8(24) & 1.541\,943\,7(24) & -0.002\,86(15)  & 1.539\,09(15) \\
	30.0 & 1.393\,761\,45(5) & 0.146\,451\,2(26) & 1.540\,212\,6(26) & -0.002\,21(12)  & 1.538\,00(12) \\
	32.5 & 1.391\,067\,28(5) & 0.148\,290\,4(28) & 1.539\,357\,7(28) & -0.001\,711(97) & 1.537\,646(97) \\
	35.0 & 1.389\,047\,31(5) & 0.150\,069\,3(30) & 1.539\,116\,6(30) & -0.001\,324(79) & 1.537\,792(79) \\
	37.5 & 1.387\,530\,09(5) & 0.151\,780\,1(32) & 1.539\,310\,2(32) & -0.001\,025(64) & 1.538\,285(64) \\
	40.0 & 1.386\,391\,69(5) & 0.153\,422\,3(34) & 1.539\,814\,0(34) & -0.000\,794(51) & 1.539\,020(51) \\
	42.5 & 1.385\,542\,54(5) & 0.154\,995\,7(35) & 1.540\,538\,2(35) & -0.000\,614(41) & 1.539\,924(41) \\
	45.0 & 1.384\,916\,73(4) & 0.156\,503\,7(37) & 1.541\,420\,4(37) & -0.000\,476(33) & 1.540\,945(33) \\
	47.5 & 1.384\,464\,91(3) & 0.157\,947\,4(39) & 1.542\,412\,3(39) & -0.000\,368(27) & 1.542\,044(27) \\
	50.0 & 1.384\,149\,48(3) & 0.159\,332\,5(41) & 1.543\,482\,0(41) & -0.000\,285(21) & 1.543\,197(21) \\
	52.5 & 1.383\,941\,70(2) & 0.160\,660\,9(44) & 1.544\,602\,6(44) & -0.000\,221(17) & 1.544\,382(18) \\
	55.0 & 1.383\,819\,27(2) & 0.161\,931\,6(45) & 1.545\,750\,9(45) & -0.000\,171(14) & 1.545\,580(15) \\
	57.5 & 1.383\,764\,60(2) & 0.163\,109(31)    & 1.546\,874(31)  & -0.000\,132(11) & 1.546\,742(33) \\
  \end{tabular}
\end{ruledtabular}
\end{table*}

It is crucial to properly choose the $\Omega$ parameters of the basis, in order to obtain sufficiently accurate exchange energy.   
If we assume $\Omega_B=\Omega_C=0$, i.e. we allow only for a nontrivial dependence in $\eta_1$ and $\eta_2$, the energy splitting
has a very low numerical convergence in $\Omega_A$, and thus this shell parameter has to be sufficiently
large to saturate the splitting.
An essential feature of exponential function basis, other than the desired analytical behavior, is its exponential convergence
to the complete basis set (CBS) limit, i.e. the $\log$ of differences of energies calculated for subsequent values of $\Omega$
are very well fitted to the linear function. By virtue of this property, extrapolation to the CBS limit is straightforward
and reliable. Interestingly, we observe analogous behavior for the energy splittings, which entails,
\begin{equation}
\frac{\Delta E\left(\Omega_{A}\right) - \Delta E\left(\Omega_{A}-1\right)}{\Delta E\left(\Omega_{A} - 1 \right) - \Delta E\left(\Omega_{A} - 2\right)} = {\rm const}\,.
\end{equation}
Consequently, extrapolation to the CBS limit can be performed by linear regression of the logarithm of energy splitting increments,
\begin{equation}
\log\left(\Delta E\left(\Omega_A\right) - \Delta E\left(\Omega_A - 1\right) \right) = \alpha - \beta~\Omega_A.
\end{equation}
The obtained coefficient $\beta$ varies from $0.121(2)$ for~$R=20.0$,~$\Omega_A = 40$ down to $0.0366(2)$ for $R=57.5$,~$\Omega_A = 145$.

At the largest considered nuclear distance of 57.5 au saturation was achieved with $\Omega_A$ as large as $145$.
This explains why previous numerical attempts were not successful. In fact,
the only correct results were obtained in a previous work by one of us (KP) in Ref.~\cite{bo_h2}, but those calculations
were performed only for internuclear distances up to $R=20$ au.

In contrast, numerical convergence in $\Omega_C$ is very fast.
We performed calculations with increasing values of $\Omega_A$ and $\Omega_C$ shell
parameters, but with $\Omega_B=0$ up to $R=57.5$. 
At first, saturation is achieved in the $\Omega_A$ parameter, and subsequently $\Omega_C$ is raised. 
Although such a basis has a multiplicative structure, a value as small as $\Omega_C=4$ was
sufficient to achieve the claimed numerical precision. 
Extrapolation is analogous as described above for the basis with $\Omega_C = 0$.
Corresponding numerical results for bases with saturation in $\Omega_A$ and $\Omega_C$ are presented in the fourth column of Table \ref{table1}.
The only exception is the case of $R=57.5$ au, for which only $\Omega_C=3$ was technically feasible,
which is reflected in higher uncertainty due to extrapolation in the $\Omega_C$ parameter.
The limiting factor was the available computer memory of 2TB, which was exhausted by recursive derivation
of integrals with extended-precision arithmetic.

The numerical convergence in $\Omega_B$ is relatively slow, but the crucial point is that the numerical
significance of the basis functions with $n_4+n_5>0$ becomes exponentially small in the limit of large internuclear distance $R$.
Therefore, we calculate $\delta = R^{-5/2}e^{2R} \left[\Delta E - \Delta E\left(\Omega_B=0\right)\right] $ using the single shell
parameter $\Omega$ as in Eq.\,(\ref{shell}) for all values of internuclear distances up to $R=35$,
which was the upper limit set by the available computer memory.
The resulting $\delta$, as a function of $R$, is very well fitted to the exponential functions of the form $\alpha\,e^{-\beta\,R}$
with $\alpha = -0.0477(8)$ and $\beta=0.1025(9)$,
and we use this fit to obtain extrapolated $\delta$ for internuclear distances $R>35$ au, as shown in Table I.
This demonstrates that the correct asymptotics of the exchange energy can be obtained using basis functions
with $n_4+n_5=0$ only. Nevertheless, the influence of functions with $n_4+n_5>0$ is included in order to obtain a complete
numerical result for the exchange energy at individual values of $R$.

The final results for the splitting, presented in the last column of Table \ref{table1},
are obtained as a sum of $\Delta E(\Omega_B=0)$ and $\delta$. 
In view of the limitations of the available computer memory and reasonable
computation times, we were able to perform calculations in $\Omega_B=0$ basis for $R$ up to $57.5$ au.
To illustrate the computational cost of these calculations, at $R=57.5$ au, the largest internuclear distance presented, $\Omega_A =
145$, was required for saturation, which amounts to solving a dense eigenproblem in approximately 230 decimal-digit precision with
basis size $N\sim27500$.

\section{Numerical fit of the asymptotic expansion}
\begin{figure}[h]
\centering
\includegraphics[width=\columnwidth]{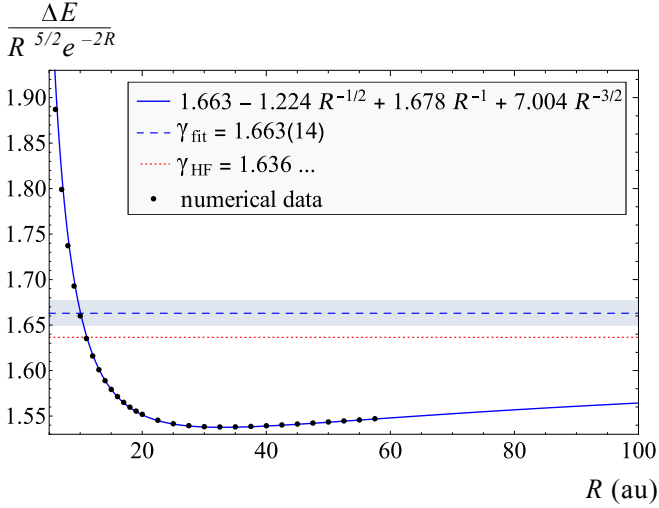}
\caption{\label{figure1} Rescaled energy splitting $R^{-5/2}\,e^{2\,R}\,\Delta E$,
fitted to numerical points in Table \ref{table1} in the range $R=20 - 57.5$ au, 
Herring and Flicker~\cite{herring} asymptotics is the red horizontal dotted line, fitted $\gamma$ is blue dashed line
and light blue shaded region represents values within the uncertainty $\sigma = 0.014$ of $\gamma$.
Results for $R\in(6,19)$ au are taken from Ref.~\cite{bo_h2} for completeness but were not used for fitting.
Values of higher order coefficients are presented without any uncertainties
because they strongly depend on the length of the expansion.
They are shown to represent the actual fitting function.}
\end{figure}
\begin{figure}[h]
\centering
\includegraphics[width=\columnwidth]{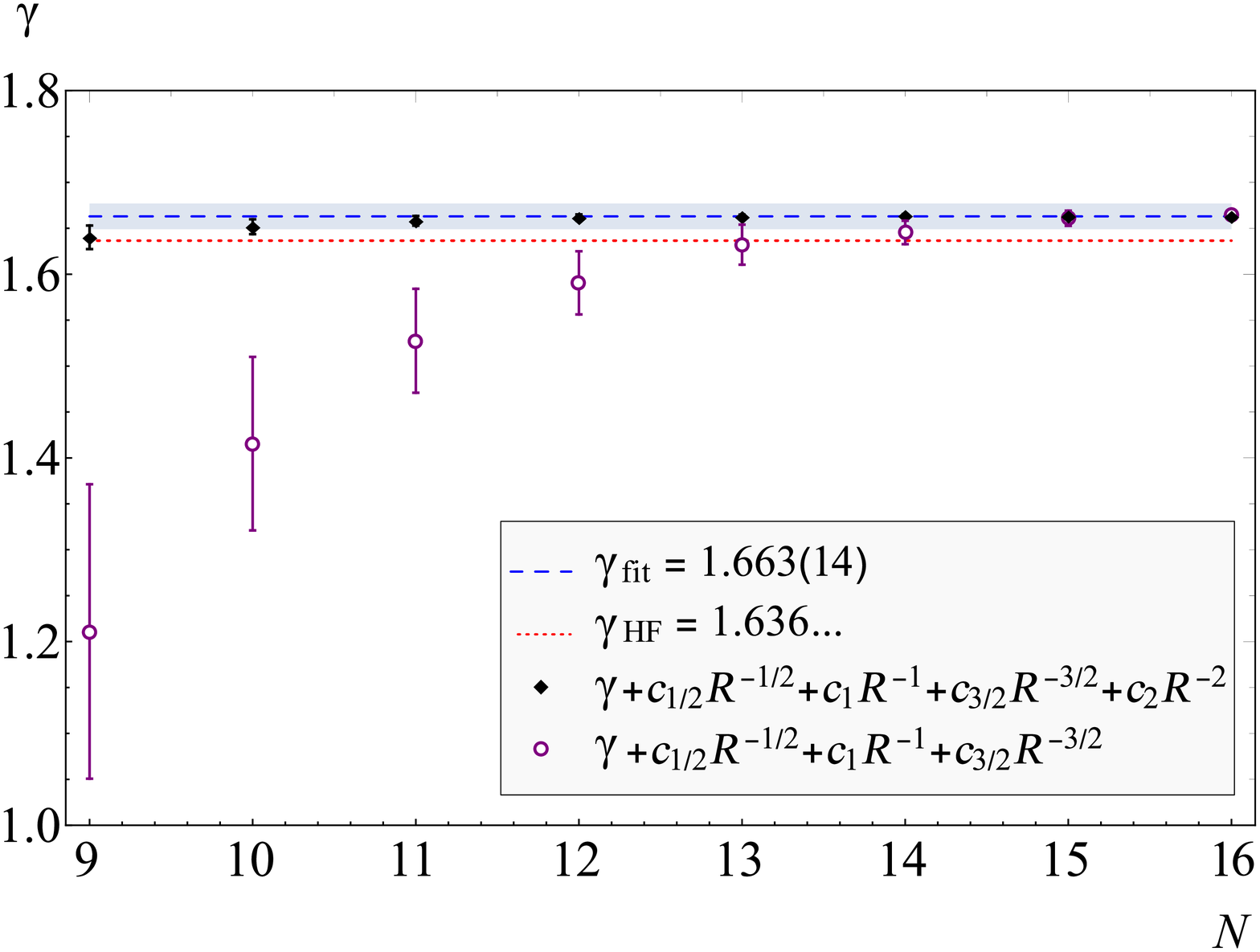}
\caption{\label{figure2} Dependence of the leading coefficient in a fit to the rescaled energy splitting
$R^{-5/2}\,e^{2\,R}\,\Delta E$, as a function of the number of last numerical data points used for fitting.
Error bars represent standard deviation of the leading coefficient resulting from linear regression.}
\end{figure}

A brief analysis of numerical results gathered in Table~\ref{table1} and depicted in Fig.~\ref{figure1}
reveals that even after rescaling by a factor of $R^{-5/2}e^{2R}$ numerical data are still distant from $\gamma$
in Eq. (\ref{gammaHF}).
Nevertheless, around $R=35$ au monotonicity changes and convergence
of $R^{-5/2}\,e^{2\,R}\, \Delta E(R)$ to the constant $\gamma$ can be observed.
In order to provide comprehensive analysis by performing a numerical fit,
an important conclusion from the Herring-Flicker work~\cite{herring} should be
recalled, i.e. that the next asymptotic term should be of the relative order $1/\sqrt{R}$, not $1/R$.
This conclusion is also supported by the derivation of Gor'kov and Pitaevskii presented in Sec. III.
The considered relatively wide region of internuclear distances enforces accounting for at least 3 or 4 terms of
asymptotic series to properly model the observed dependence.
The result for the leading term depends on the length of the fitting expansion, but converges
to the same value with an increasing number of points used for fitting, see Fig. \ref{figure2}.
Taking this into account, and the number of terms in the asymptotic series,
we can estimate the leading coefficient to be $1.663(14)$, which is $2\sigma$ away from the Herring-Flicker value
in Eq. (\ref{gammaHF}), where we use the convention that the number in parentheses is the uncertainty
denoted in the text by $\sigma$. This uncertainty is obtained by studying fits of various lengths
to variable numbers of points and is chosen very conservatively.
\noindent
Because  the original calculations in Refs. \cite{herring,gorkov} lack mathematical rigour and the asymptotic wave
function is non-differentiable at $z_1 = -z_2$, the value of the asymptotics might be not fully correct.

Nonetheless, by assuming correctness of the Herring-Flicker value, which amounts
to fixing $\gamma = \gamma_{\rm HF}$, and subsequent fitting in powers of $1/\sqrt{R}$,
a coefficient for the next-to-leading, $R^{2}e^{-2R}$, term can be estimated as $-0.66(7)$,
which is significant, as conjectured by Hirschfelder and Meath~\cite{intermol}.
This value is in disagreement with the work of Andreev~\cite{andreev},
in which the next non-vanishing term is claimed to be $R^{3/2}\,e^{-2\,R}$. 
\begin{figure}[h]
\centering
\includegraphics[width=\columnwidth]{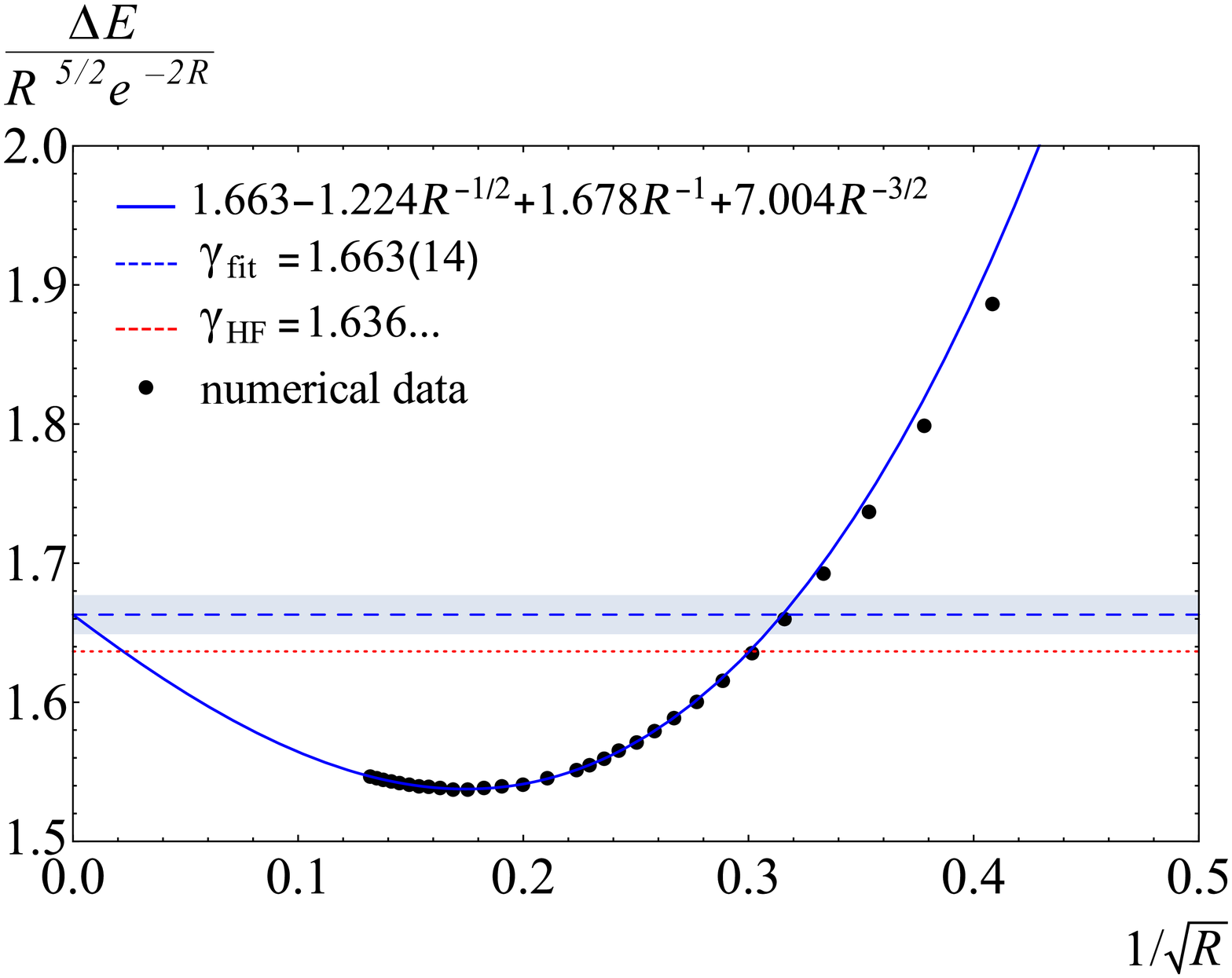}
\caption{\label{figure3} Rescaled energy splitting $R^{-5/2}\,e^{2\,R}\,\Delta E$, 
  fitted to numerical points in Table \ref{table1} in the range $R=20 - 57.5$ au,
  the same as in Fig.~\ref{figure1}, but represented as a function of $1/\sqrt{R}$.}
\end{figure}

It is perhaps more convenient to present numerical results and the fit as a function of $1/\sqrt{R}$, see Fig. \ref{figure3}.
Then it becomes more evident, that the calculated numerical values are sufficient to obtain the leading coefficient, 
and the polynomial fit should consist of at least 3 or 4 terms to properly model the numerical data.

Curiously, in the aforementioned $\Omega_B=\Omega_C=0$ basis, the rescaled exchange splitting $R^{-5/2}e^{2R} \Delta E$ 
quickly and monotonically converges as a function of $R$, to a constant value of $\gamma_0=1.3835(2)$.
Therefore, even in a basis with no powers of $r_{12}$, leading asymptotic behavior could be achieved, although
with the slightly smaller coefficient $\gamma_0$.
Inclusion of higher powers of $r_{12}$ brings
this constant close to $\gamma_{\rm HF}$, but even for the largest distances considered in the calculations,
numerical points are still distant from the asymptotic constant, as presented in Fig.\,\ref{figure1}.

\begin{figure}[h]
\centering
\includegraphics[width=\columnwidth]{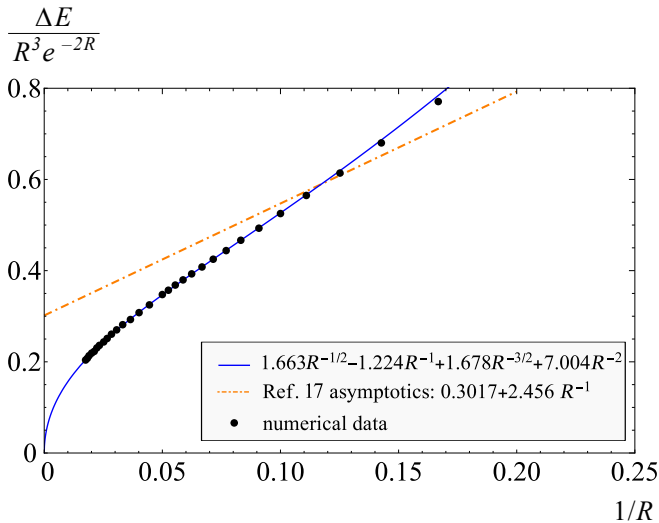}  
\caption{\label{figure4}  Rescaled energy splitting $R^{-3}\,e^{2\,R}\,\Delta E$ fitted
to numerical points in Table \ref{table1} in the range $R=20 - 57.5$ au, the same as in Fig.~\ref{figure1},
but presented as a function of $1/R$. Burrows-Dalgarno-Cohen asymptotics~\cite{burrows} is the dash-dotted orange line.}
\end{figure}

Considering the result of Ref.~\cite{burrows}, their leading asymptotics seems to be in significant disagreement
with our numerical data, see Fig. \ref{figure4}.
This asymptotics, even with inclusion of a few higher order terms, cannot match our numerical data.
Nevertheless, if one assumes the leading asymptotics of the form $R^3\,e^{-2R}$, although with the unknown coefficient,
the results of the fit of a polynomial in $1/R$ strongly depend on the length of the fitting series
and the number of points used for fitting.
The obtained coefficients are abnormally large and have an alternating sign.
This is an indication of improper choice of fitting function.
If, nevetheless, one assumes $R^3e^{-2R}$ asymptotics and fits a similar polynomial in $1/\sqrt{R}$
as in Fig. \ref{figure1} to the rescaled energy, one obtains
\begin{align}
	R^{-5/2}\,e^{2\,R}\,\Delta E =&\ 0.000\,06(83) R^{1/2} + 1.662 -1.212\,R^{-1/2} \nonumber \\ &\
  + 1.632\,R^{-1} - 7.070\,R^{-3/2}\,.
\end{align}
The leading coefficient is consistent with 0, what is in disagreement with
$\gamma_{\rm BDC}$ from Ref.~\cite{burrows}.
This disagreement is even more pronounced when numerical results are confronted with the asymptotics
of Ref. \cite{burrows} in Fig. \ref{figure4}.

\section{Conclusions}
The high numerical accuracy for the exchange energy is achieved owing not only to the correct asymptotic
behavior of explicitly correlated exponential functions, but also due to the specific choice of the basis functions
suggested by the significance of the internuclear axis neighborhood.
Regardless of the relatively limited range of internuclear distances at $R\leq 57.5$ au,
due to this high numerical accuracy, we were able to resolve the long-standing discrepancy between
long-range asymptotics. Our results are in agreement with $\gamma\,R^{5/2}\,e^{-2\,R}$ asymptotics,
although our numerically fitted $\gamma$ is $2\sigma$ away from the Herring-Flicker value $\gamma_{\rm HF}$.
Notably, fits of different lengths converge to the same $\gamma$, as shown in Fig. \ref{figure2}, while
the fit to $R^3\,e^{-2\,R}$ asymptotics gives a very small coefficient, consistent with $0$ and
in strong disagreement with $\gamma_{\rm BDC}$. This disagreement becomes more evident
when the asymptotics from Ref.~\cite{burrows} is confronted with our numerical results in Fig. \ref{figure4}.

To conclude, our numerical results revise the recent analytic derivations of the large-distance asymptotics
and will provide a valuable benchmark for various calculations of interatomic interactions.

\section*{Acknowledgments}

The authors wish to acknowledge valuable discussions with Bogumi\l\ Jeziorski and are grateful to Heiko\ Appel for
providing access to large memory computer resources.
This work was supported by the National Science Center (Poland) Grant No. 2017/27/B/ST2/02459.

\section*{Data Availability}
Detailed numerical data are available on request from the corresponding author (MS).

\end{document}